\input{aipcheck}

\documentclass[
    ,final            
  ]
  {aipproc}

\layoutstyle{6x9}

\usepackage{epstopdf}
\newcommand{\beq}{\begin{equation}}
\newcommand{\eeq}{\end{equation}}
\newcommand{\bea}{\vspace{0.25cm}\begin{eqnarray}}
\newcommand{\eea}{\end{eqnarray}}

\newcommand{\qb}{\mbox{{\bf
q}}}

\newcommand{\rb}{\mbox{{\bf
r}}}

\def\lsim{\mathrel{\rlap{\lower4pt\hbox{\hskip1pt$\sim$}}
    \raise1pt\hbox{$<$}}}         
\def\gsim{\mathrel{\rlap{\lower4pt\hbox{\hskip1pt$\sim$}}
    \raise1pt\hbox{$>$}}}         

\begin{document}

\title{
Nuclear higher-twist effects in eA DIS}



\author{B.G.~Zakharov}{
address={L.D. Landau Institute for Theoretical Physics,
        GSP-1, 117940,\\ Kosygina Str. 2, 117334 Moscow, Russia
}
}

\begin{abstract}
We discuss the relation between the treatments
of the higher twist nuclear effects in $eA$ DIS based on the 
pQCD collinear approximation and the light-cone path
integral formalism. 
We show that in the collinear approximation the $N\!=\!1$ 
rescattering contribution to the gluon emission vanishes. 
It is demonstrated that the nonzero gluon spectrum obtained 
by Guo, Wang and Zhang is a consequence of unjustified neglect of some terms
in the collinear expansion.
\end{abstract}

\maketitle


\noindent{\bf 1}.
In this talk, I focus on the relation between 
the treatment of the induced gluon emission from the fast struck quark
in $eA$ DIS in the higher-twist pQCD approach
\cite{W1,W2} 
and in the light-cone path integral (LCPI) approach \cite{LCPI}. 
The first approach uses the Feynman diagram technique
and the collinear approximation
(for a review, see \cite{QS}).
The LCPI formalism is formulated in terms of the wave functions in
the coordinate space.
At the end of the talk I discuss the physical interpretation of 
the $1/Q^{2}$ suppression of soft nuclear rescatterings in jet production
without the induced gluon emission.

\noindent{\bf 2}.
In \cite{W1,W2} the twist-4 contribution ($N\!=\!1$ rescattering)
to the gluon spectrum has been expressed in terms of the diagrams 
like shown in Fig.~1.
\begin{figure}[ht]
\includegraphics[height=.165\textheight]{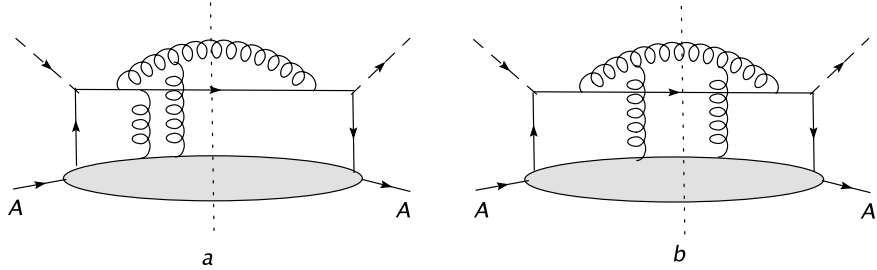}
\caption{Examples of the diagrams for gluon emission in the higher-twist 
approach \cite{W1,W2}.}
\end{figure}
The lower soft parts of the diagrams are expressed via the matrix element
$
\langle A|\bar{\psi}(0)A^{+}(y_{1})A^{+}(y_{2})\psi(y_{3})|A\rangle\,.
$
The upper hard parts, $H$, are calculated perturbatively.
The contour integrations over the $p^{+}$ momentum components are carried out
with the help of the poles in the denominators of the retarded 
propagators for fast quarks and gluon 
\beq
G_{ret}(y)=\frac{i}{(2\pi)^{4}}\int\limits_{ {p^{-}> 0}} 
dp^{+}dp^{-}d\vec{p}_{T}
\frac{\exp{[-i(p^{+}y^{-}+p^{-}y^{+}-\vec{p}_{T}\vec{y}_{T})]}}
{2p^{-}\left(p^{+}-\frac{\vec{p}_{T}^{2}+m^{2}-i0}{2p^{-}}\right)
}\,
\label{eq:10}
\eeq
($p^{\pm}=(p^{0}\pm p^{3})/\sqrt{2}$,
the virtual photon momentum is chosen in the negative $z$ direction,
for simplicity spin is ignored).
The collinear approximation corresponds to replacement of the hard part by
its second order expansion in the $t$-channel transverse gluon 
momentum $\vec{k}_{T}$ (only the second order term is important)
\beq
H(\vec{k}_{T})\approx
H(\vec{k}_{T}=0)+
\left.\frac{\partial H}
{\partial k^{\alpha}_{T}}\right|_{\vec{k}_{T}=0}k^{\alpha}_{T}
+
\left.\frac{\partial^{2} H}
{\partial k^{\alpha}_{T} \partial k^{\beta}_{T}}
\right|_{\vec{k}_{T}=0}\cdot
\frac{k^{\alpha}_{T}k^{\beta}_{T}}{2}
\,.
\label{eq:20}
\eeq

\noindent{\bf 3}.
In the LCPI approach the amplitude for $a\rightarrow bc$ partonic
process is written in terms  of the incoming ($i=a$) and 
outgoing ($i=b,c$) wave functions. These functions are written as 
$
\psi_{i}(y)=\frac{1}{\sqrt{2p^{-}_{i}}}\exp[-ip^{-}_{i}y^{+}]
 \phi_{i}(y^{-},\vec{y}_{T})\,.
$
The $y^{-}$ dependence of the transverse wave functions
$\phi_{i}$ is governed by the two-dimensional 
Schr\"odinger equation
\beq
i\frac{\partial\phi_{i}(y^{-},\vec{y}_{T})}{\partial
y^{-}}=
\left\{\frac{
[(\vec{p}_{T}-g\vec{A}_{T})^{2}
+m^{2}_{i}]} {2 p_{i}^{-}}
+gA^{+}\right\}
\phi_{i}(y^{-},\vec{y}_{T})\,.
\label{eq:30}
\eeq
Diagrammatically the gluon emission from the struck quark 
can be described by the graph shown in Fig.~2, where 
the horizontal lines show the Green's functions 
${\cal{K}}$ ($\rightarrow$) and 
${\cal K}^{*}$ ($\leftarrow$) of the equation (\ref{eq:30}).
\begin{figure}[!htb]
   \centering
\includegraphics[height=.13\textheight]{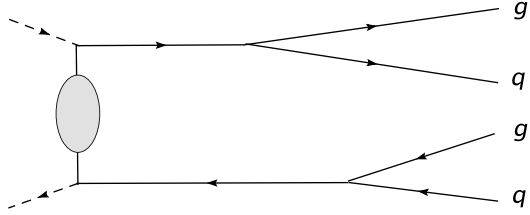}
\caption{Schematic diagram for gluon emission in the LCPI approach 
\cite{LCPI}.}
\end{figure}
The integration over { $y^{+}$} of the frequently oscillating factors
can be done in the same way as for the ordinary quark distribution $f_{q}$.
As compared to the Collins-Soper formula  
{ $f_{q}=\frac{1}{4\pi}\int dy^{-}e^{ix_{B}P^{+}y^{-}}
\langle N|\bar{\psi}(-y^{-}/2)\gamma^{+}\psi(y^{-}/2)|N\rangle$}, now
$e^{ix_{B}P^{+}y^{-}}$ is replaced by 
$$e^{ix_{B}P^{+}y^{-}}\!\int d\xi d\xi'
{\cal{K}}_{q_{f}}(\infty|\xi){\cal{K}}_{g}(\infty|\xi)
{\cal{K}}_{q_{i}}(\xi|y^{-}/2)
{\cal{K}}_{q_{f}}^{*}(\infty|\xi')
{\cal{K}}_{g}^{*}(\infty|\xi'){\cal{K}}_{q_{i}}^{*}(\xi'|-y^{-}/2)$$
(for simplicity we omit all the transverse coordinates).
One can show that the LCPI formalism can be obtained from
the pQCD treatment in terms of Feynman diagrams. Indeed,
the retarded propagator (\ref{eq:10}) 
can be written in terms of the zeroth-order (in $A_{\mu}$) Green's function of 
the Schr\"odinger equation (\ref{eq:30})
(we denote it $K$) as
\beq
G_{ret}(y)=\int_{0}^{\infty}\frac{dk^{-}}{4\pi k^{-}}
e^{-ik^{-}y^{+}}
K(\vec{y}_{T},y^{-}|0,0)\,.
\label{eq:40}
\eeq
The integration in the Feynman diagrams over $y^{+}$ gives conservations
of the large $k^{-}$ components in each vertex, and we reproduce
exactly the LCPI expression in terms of  the transverse Green's functions.

\noindent{\bf 4}.
The typical difference in the coordinate $y^{-}$ for the 
upper and lower $\gamma^{*}qq$ vertices 
in Fig.~2
(which gives the scale of the 
quantum nonlocality of the fast quark production)
is given by the well known Ioffe
length $L_{I}=1/m_{N}x_{B}$. For moderate values of $x_{B}$ this 
scale is much smaller than the typical scale of the gluon emission.
It allows one to treat the quark production and gluon emission 
as mutually independent. Then for a given position $\rb$ of the struck nucleon 
the generalized quark distribution as a function 
of the gluon fractional momentum $z$ can be approximated by the 
factorized form 
\beq
{df_{q}(z,\rb)}/{dz}\approx f_{q}\,{dP(z,\rb)}/{dz}\,,
\label{eq:50}
\eeq
where the quark distribution $f_{q}$ stems from the left part of the diagram
in Fig.~2, and the gluon spectrum $dP/dz$ is described by the right parts of 
the diagram evaluated neglecting the quantum nonlocality of the 
fast quark production. 

In the LCPI approach \cite{LCPI} the final formula for $dP/dz$ is expressed 
in terms of the Green's function for the two-dimensional 
Schr\"odinger equation with an imaginary potential which is proportional 
to the dipole cross section $\sigma(\rho)$. 
The Hamiltonian takes the harmonic oscillator form
for a quadratic parametrization $\sigma(\rho) = C\, \rho^2$.
The $N\!=\!1$ contribution to the gluon spectrum in the oscillator 
approximation should coincide
with prediction of the collinear approximation in the treatment \cite{W1,W2}.
Indeed, the quadratic form of the dipole cross section 
corresponds to the vector potential 
approximated by the linear expansion 
$
A^{+}(y^{-},\vec{y}_{T}+\vec{\rho})\!\approx\!
A^{+}(y^{-},\vec{y}_{T})+
\vec{\rho}
\nabla_{y_{T}}
A^{+}(y^{-},\vec{y}_{T})\,
$
which can be traced back to the collinear expansion in the momentum space.
For massless partons  in the oscillator approximation
the spectrum of the LCPI approach coincides with the
spectrum in the BDMPS formalism \cite{BDMPS}. For a target of thickness $L$
it reads \cite{BDMPS,Z_OA} 
\beq
\frac{dP}{dz}=\frac{\alpha_{s}P_{Gq}(z)}{\pi}
\ln|\cos\Omega L|\,,
\label{eq:60}
\eeq
where $P_{Gq}$ is the ordinary splitting function, 
$\Omega=\sqrt{-i{C_{3} n}/{z(1-z)E}}$, $C_{3}=C C_{A}/C_{F}$.
The expansion of the spectrum (\ref{eq:60}) in the density 
starts with the terms $\propto n^{2}$ which corresponds to the $N\!=\!2$ 
rescatterings \cite{Z_OA}, and
the contribution of $N\!=\!1$ rescattering is absent. 
The fact that the $N\!=\!1$ gluon spectrum vanishes in the collinear
approximation  can also be demonstrated by the direct calculations 
of the second derivative of the hard part in the momentum space 
\cite{Z_OA,AZZ}.

\noindent{\bf 5}.
Let us now discuss why a nonzero $N=1$ spectrum has been obtained 
in \cite{W1,W2}. In \cite{W1,W2} the nonzero second derivative
of the hard part (at $z\ll 1$) comes from the graph shown in 
Fig.~1b (at $z\ll 1$).
The authors use for the integration variable in the hard part of
this graph the transverse momentum of the final gluon, $\vec{l}_{T}$.
The $\vec{l}_{T}$-integrated
hard part obtained in \cite{W2} (Eq. 15 of \cite{W2}) reads
(up to an unimportant factor)  
\beq
H(\vec{k}_{T})\propto \int \frac{d\vec{l}_{T}}
{(\vec{l}_{T}-\vec{k}_{T})^{2}} R(y^{-},y^{-}_{1},y^{-}_{2},
\vec{l}_{T},\vec{k}_{T})\,,
\label{eq:70}
\eeq
\bea
{R}(y^{-},y^{-}_{1},y^{-}_{2},
\vec{l}_{T},\vec{k}_{T})=\frac{1}{2}
\exp{
\left[i
\frac
{y^{-}(\vec{l}_{T}-\vec{k}_{T})^{2}-
(1-z)(y^{-}_{1}-y^{-}_{2})(\vec{k}_{T}^{\,2}
-2\vec{l}_{T}\vec{k}_{T})}{2q^{-}z(1-z)}
\right]
}\nonumber\\
\times\left[
1-\exp{
\left(i
\frac{
(y^{-}_{1}-y^{-})(\vec{l}_{T}-\vec{k}_{T})^{2}}
{2q^{-}z(1-z)}\right)}
\right]
\cdot
\left[
1-\exp{
\left(-i
\frac{
y^{-}_{2}(\vec{l}_{T}-\vec{k}_{T})^{2}}
{2q^{-}z(1-z)}\right)}
\right]
\label{eq:80}
\eea
($y^{-}$, $y^{-}_{1,2}$ correspond to the quark interactions 
with the virtual photon and $t$-channel gluons,
our $z$ equals $1-z$ in \cite{W1,W2}).
The dominating nuclear-size enhanced configurations correspond to 
$|y^{-}_{1}-y^{-}_{2}|\lsim R_{N}$, 
$|y^{-}|\lsim R_{N}$ ($R_{N}$ is the nucleon size). 
This contribution can be calculated taking $y^{-}_{1}=y^{-}_{2}$. 
A simple calculation gives
\beq 
\nabla_{k_{T}}^{2}H|_{\vec{k}_{T}=0}
\propto
4\pi \int\limits_{0}^{\infty}
dl_{T}^{2}
\left\{
\frac{1-\cos(a l_{T}^{2})}{l_{T}^{4}}
-
\frac{a\sin(a l_{T}^{2})}{l_{T}^{2}}
+
a^{2}\cos(a l_{T}^{2})
\right\}\,,
\label{eq:90}
\eeq
where
$a=y_{1}/2q^{-}z(1-z)$. It gives $\nabla_{k_{T}}^{2}H|_{\vec{k}_{T}=0}=0$. Indeed,
after integrating the first term by parts
it cancels the contribution from the second term in (\ref{eq:90}).
The last integral equals zero. 
The last two terms in the integrand in (\ref{eq:90}) which come from
differentiating the factor $R$ have been unjustifiedly neglected
in \cite{W1,W2}. It is for this reason that the authors have obtained
the  nonzero spectrum. 

\noindent{\bf 6}.
The pQCD calculations within the collinear approximation 
\cite{QS} show that the $N=1$ rescattering 
contribution in jet production (in $eA$ DIS or in $hA$ collisions) 
is $\propto A^{1/3}\times \lambda /Q^{2}$, where
$\lambda$ can be expressed in terms the higher-twist parton distribution.
The fact that the additional soft rescattering is suppressed
by $ 1/Q^{2}$ is surprising. In \cite{QS} this fact was attributed to
the off-shell suppression of the rescattering cross section for virtual
partons.
However, this interpretation is inconsistent with the uncertainty 
relation $\Delta E \Delta t\gsim 1$. Let us consider jet production
in $hA$-collisions. In this case from the uncertainty relation one 
can obtain for the $L$-dependence of the fast parton virtuality 
$Q^{2}(L)\sim Q^{2}_{in}/Lm$ ($m\sim m_{N}/3$, $Q_{in}\sim p$).
It says that the typical virtuality at $L\sim R_{A}$ (which dominates
the soft rescatterings) becomes small $\langle 1/Q^{2}\rangle \gg 1/Q^{2}_{in}$.
Note also that in pQCD calculations the distance between 
the parton production point and rescattering does not appear 
in the formula for rescattering contribution at all. 
It makes clear that $1/Q^{2}$ suppression is not due to the parton 
off-shellness.
A simple interpretation of the $1/Q^{2}$ suppression can be 
given in the LCPI approach. Jet production in a large nucleus at the 
amplitude level can be described by the diagram shown in Fig.~3a,
\begin{figure}[!htb]
\begin{tabular}{c c}
\includegraphics[height=.145\textheight]{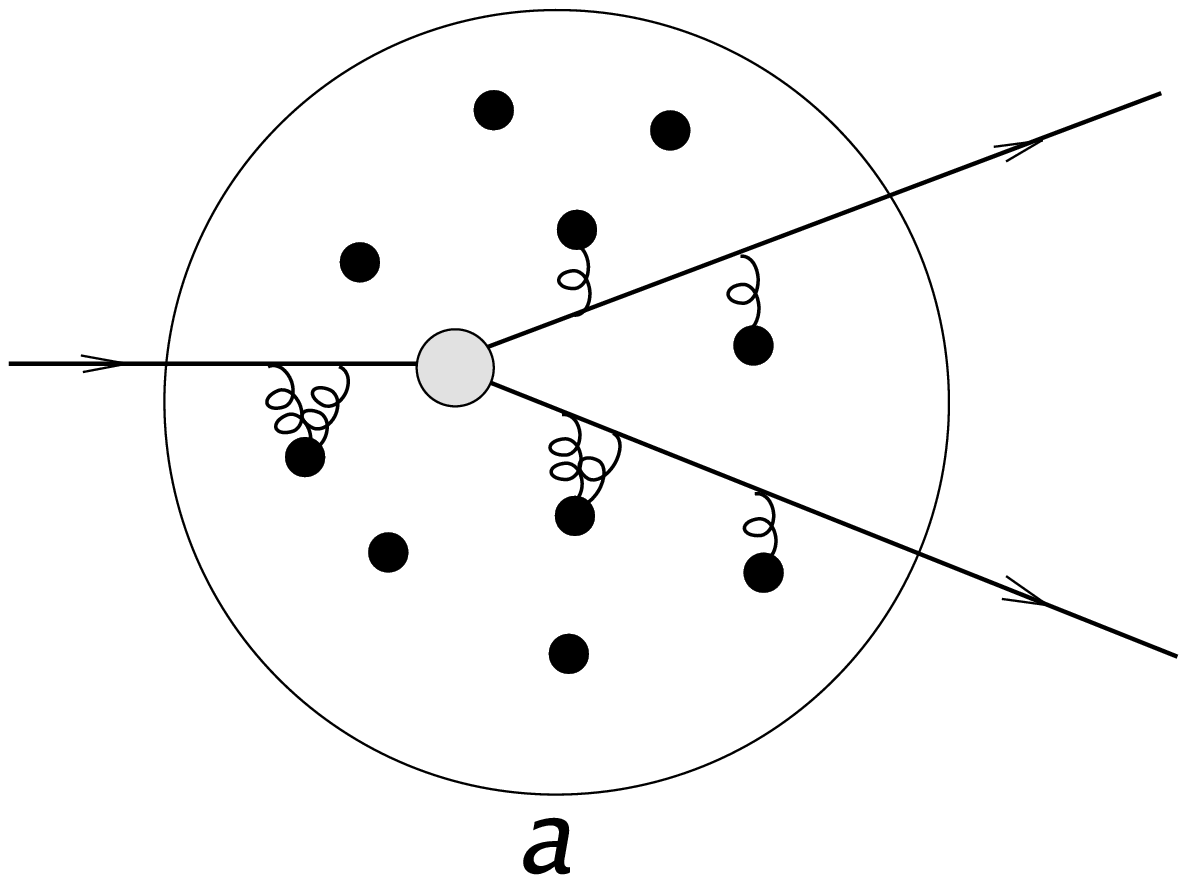}&\hspace{2cm}
\includegraphics[height=.145\textheight]{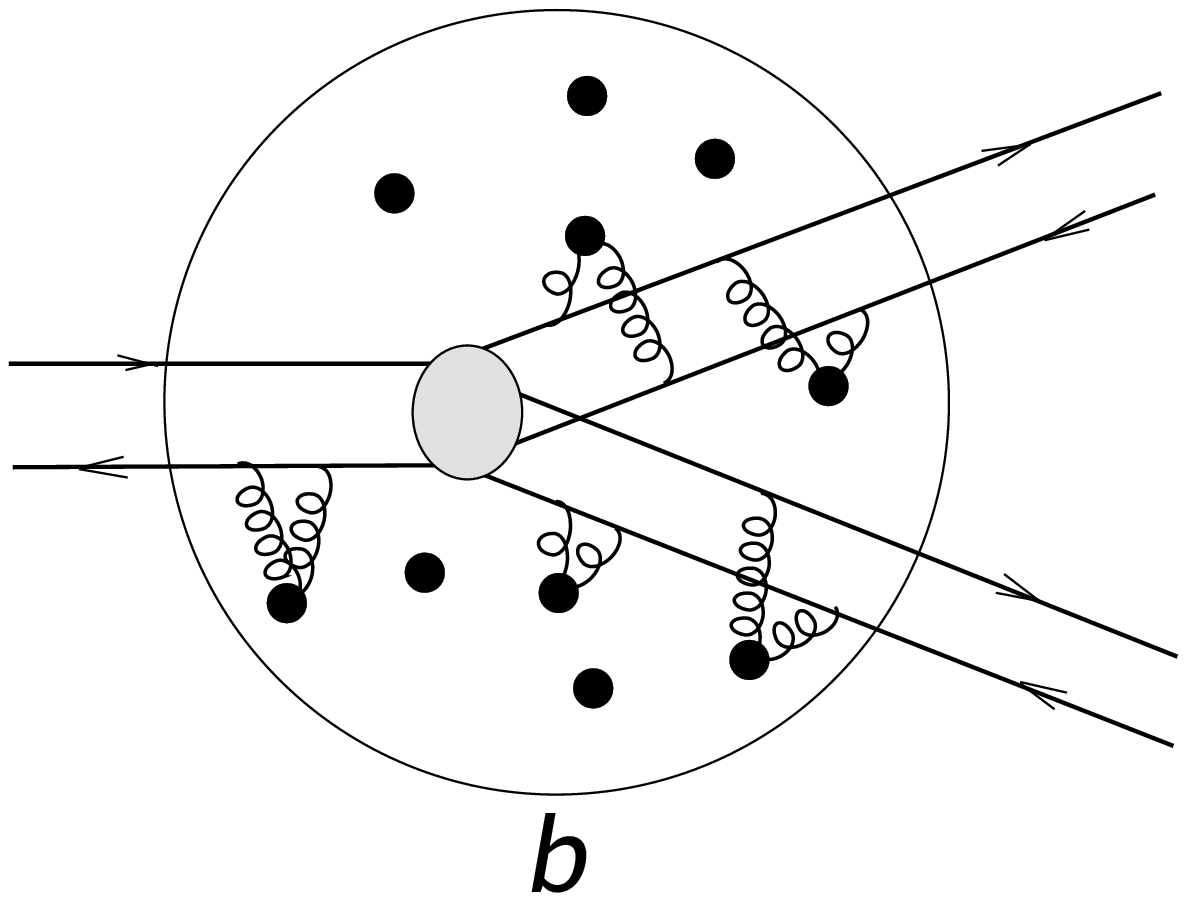}\\
\end{tabular}
\caption{Diagrams 
for the amplitude (a) and cross section (b) of jet production on a nucleus.}
\end{figure}
and the cross section is described by the diagram shown in Fig.~3b.
For each nucleon the sum of the double gluon exchange 
diagrams which appear in Fig.~3b can be written
in terms of the dipole cross section $\sigma(\rho)$.
The in-medium evolution of the ``parton-antiparton''  systems 
in Fig.~3b are
described by the evolution operator for the density matrix
which can be calculated exactly  in the path integral method 
\cite{Z1987,LCPI}. Remarkably, the path integral calculation
reproduces the result of the eikonal approximation which
neglects the transverse motion. It means that the trajectories
in Fig.~3b are essentially the straight lines. Since the transverse
distance between $\rightarrow$ and $\leftarrow$ lines is $\sim 1/p$
using $\sigma(\rho)\sim \alpha_{s}^{2}\rho^{2}xf_{g}(x,p\sim 1/\rho)$,
we obtain for the nuclear correction
$\delta\sigma^{hard}_{A}(p)/
\sigma^{hard}_{N}(p)
\sim n_{A}L xf_{g}(x,p)/p^{2}\,
$
(here $f_{g}$ is the gluon density, $x\sim 0$).
This result can be reproduced in the momentum space as well.
The $p_{T}$ broadening distribution for each parton 
reads \cite{Z1987,LCPI}
\beq
I(\vec{q})=\frac{1}{(2\pi)^{2}}\int d\vec{\rho} 
\exp{[i\vec{q}\vec{\rho}-L\sigma(\rho)n/2]}\,.
\label{eq:100}
\eeq
This $I(\vec{q})$ coincides with the prediction of the classical
transport equation evaluated with rescattering cross section 
$d\sigma/dq^{2}$ connected with $\sigma(\rho)$ by the relation
$
\sigma (\rho) = \frac{2}{\pi} \int \, d\vec{q} \, [ 1 - 
\exp (i\vec{q} \vec{\rho})]
 \frac{d\sigma}{dq^2} .
$
The in-medium hard cross section has a simple form
$$
\sigma^{hard}_{A}(\vec{p})=\int d\vec{q}
\sigma^{hard}_{N}(\vec{p}-\vec{q})I(\vec{q})
\approx
\sigma^{hard}_{N}(\vec{p})+\frac{1}{4}
\left(\frac{\partial}{\partial \vec{p}}\right)^{2}
\sigma^{hard}_{N}(\vec{p})\int_{p>q} d\vec{q} q^{2}I(\vec{q})\,
$$
which gives for the $N\!=\!1$ rescattering correction
$
\delta\sigma^{hard}_{A}(p)/
\sigma^{hard}_{N}(p)
\sim \langle q^{2}(p)\rangle /p^{2}\,\,$. 

One remark on the form of $I(\qb)$ is in order here.
The collinear expansion gives Gaussian $I(\qb)$ \cite{MM}.
However, one can easily show that for real nuclei the $I(\qb)$ calculated
accounting for the Coulomb effects turns out to be essentially 
non-Gaussian due to smallness of the nucleus size. 

\noindent {\bf 7}. 
In summary, we have demonstrated that 
the collinear expansion
gives a zero $N\!=\!1$ rescattering contribution 
to the gluon spectrum in $eA$ DIS. 
The nonzero spectrum obtained in \cite{W1,W2}
is a consequence of unjustified neglect of some terms
in the collinear expansion.
The $1/Q^{2}$ suppression of soft nuclear rescatterings is not an 
off-shell effect. The effect has a simple classical interpretation.
The collinear approximation can not be used for evaluation 
of the form of the $p_{T}$-broadening distribition for real nuclei.

\begin{theacknowledgments}
I am grateful to R.~Fiore and J.~Soffer for their hospitality at 
Diffraction 2008. I also am thankful to P.~Aurenche and H.~Zaraket
for collaboration.
This work is supported in part by the grant RFBR
06-02-16078-a and the ENS-Landau program. 
\end{theacknowledgments}

\end{document}